\documentclass[twocolumn,aps,superscriptaddress,floatfix,bibnotes,nobibnotes]{revtex4-1}

\usepackage[colorlinks=true,citecolor=blue,urlcolor=blue,linkcolor=black]{hyperref}
\usepackage{graphicx}
\usepackage{graphicx}
\usepackage{dcolumn}
\usepackage{bm}
\usepackage{lipsum}
\usepackage{amsmath}
\usepackage[mathlines]{lineno}

\begin{document}

\title[]{Ultrafast signatures of spin and orbital order in antiferromagnetic $\alpha$-Sr$_2$CrO$_4$}

\author{Min-Cheol Lee}
\affiliation{Center for Integrated Nanotechnologies, Los Alamos National Laboratory, Los Alamos, NM, USA}
\thanks{Corresponding author}
\email{mclee@lanl.gov}

\author{Connor Occhialini}
\author{Jiarui Li}
\affiliation{Department of Physics, Massachusetts Institute of Technology, Cambridge, MA, USA}

\author{Zhihai Zhu}
\affiliation{Department of Physics, Massachusetts Institute of Technology, Cambridge, MA, USA}
\affiliation{Beijing National Laboratory for Condensed Matter Physics,
Institute of Physics, Chinese Academy of Sciences, Beijing 100190, China}

\author{Nicholas S. Sirica}
\author{L. T. Mix}
\author{Dmitry A. Yarotski}
\affiliation{Center for Integrated Nanotechnologies, Los Alamos National Laboratory, Los Alamos, NM, USA}

\author{Riccardo Comin}
\affiliation{Department of Physics, Massachusetts Institute of Technology, Cambridge, MA, USA}

\author{Rohit P. Prasankumar}
\thanks{Corresponding author}
\email{rpprasan@lanl.gov}
\affiliation{Center for Integrated Nanotechnologies, Los Alamos National Laboratory, Los Alamos, NM, USA}

\date{\today}
\begin{abstract}
We used femtosecond optical spectroscopy to study ultrafast spin and orbital ordering dynamics in the antiferromagnetic Mott insulator $\alpha$-Sr$_2$CrO$_4$. This chromate system possesses multiple spin and orbital ordered phases, and therefore could enable us to study the unique interplay between these collective phases through their non-equilibrium response to photoexcitation. Here, by varying the pump photon energy, we selectively drove inter-site spin hopping between neighboring Cr $t_{2g}$ orbitals and charge transfer-type transitions between oxygen 2$p$ and Cr $e_{g}$ orbitals. The resulting transient reflectivity dynamics revealed temperature-dependent anomalies across the N${\textrm{\'e}}$el temperature for spin ordering as well as the transition temperatures linked to different types of orbital order. Our results reveal distinct relaxation timescales for spin and orbital orders in $\alpha$-Sr$_2$CrO$_4$ and provide experimental evidence for the phase transition at $T_\textrm{O}$, possibly related to antiferro-type orbital ordering.
\end{abstract}

\maketitle

 Transition metal (TM) compounds provide a versatile platform for exploring a wide variety of strongly correlated phenomena, such as high-$T_\textrm{C}$ superconductivity, multiferroicity, or novel magnetism \cite{Khomskii2014,Streltsov2017,Imada1998}. These phenomena are accompanied by various types of charge, spin and orbital orders, arising from competition between interactions such as the local crystal field, inter-site exchange interactions, spin-orbit coupling and on-site Coulomb repulsion \cite{Imada1998}. Several phase transitions can occur in a single compound at different temperatures, as observed in multi-orbital systems with partially filled $d$-shells, including cuprates \cite{Caciuffo2002}, manganites \cite{Murakami1998,Rao2000}, and ruthenates \cite{Zegkinoglou2005}. 

 Chromates are an intriguing member of this class of materials, due to their active degrees of freedom (DOFs), multiple electronic symmetry breaking pathways, and competing order parameters. The case of $\alpha$-Sr$_2$CrO$_4$ is particularly interesting, as it has been shown to exhibit multiple spin and orbital ordering transitions in the Mott insulating ground state (Fig. 1(a)) \cite{Zhu2019,Jeanneau2019,Sakurai2014,Matsuno2005,Pandey2021}. $\alpha$-Sr$_2$CrO$_4$ presents a unique electronic configuration, with two electrons in $t_{2g}$ orbitals and total spin of $S=1$. Recent resonant X-ray scattering (RXS) \cite{Zhu2019} and neutron powder diffraction studies \cite{Jeanneau2019} demonstrated antiferromagnetic (AFM) order by tracking a magnetic Bragg peak below the N\textrm{\'e}el temperature ($T_\textrm{N} = 112$ K), consistent with the temperature ($T$)-dependent magnetic susceptibility \cite{Sakurai2014}. The RXS study also revealed another in-plane order below $T_\textrm{S} = 50$ K \cite{Zhu2019}, which was attributed to stripe-type spin-orbital order in the $d_{yz}$/$d_{zx}$ orbital sub-manifold (Fig. 1(a)). $\alpha$-Sr$_2$CrO$_4$ also manifests a signature of a broad transition around $T_\textrm{O} = 140$ K in the specific heat, which is likely unrelated to spin order, as the magnetic susceptibility follows a typical Curie-Weiss law across $T_\textrm{O}$ \cite{Sakurai2014}. The exact origin of this feature has not been unveiled yet, though it was suggested to arise from a distinct antiferro(AF)-type orbital order in the $d_{yz}$ and $d_{zx}$ orbitals (Fig. 1(a)), as in Sr\textsubscript{2}VO\textsubscript{4} \cite{Sakurai2014,Zhou2007}. Therefore, more experimental insight is needed to unravel the origin of this phase transition at $T_\textrm{O}$ above the N$\textrm{\'e}$el temperature.
 
 In this context, ultrafast optical spectroscopy (UOS) is a powerful tool for providing new insight into the intertwined spin and orbital DOFs in TM compounds \cite{Giannetti2016}. By using femtosecond (fs) photoexcitation to drive the system out of equilibrium, the interplay among the competing charge, spin and orbital DOFs can be disentangled via their different timescales for returning to the initial ground state, as exemplified in previous studies of various TM oxides \cite{Fausti2011,Iwai2003,Li2013,Prasankumar2007,Bowlan2016,Prasankumar2005,Lee2019_1,Coslovich2017}. However, to the best of our knowledge this technique has not yet been used to study $\alpha$-Sr$_2$CrO$_4$, providing an exciting opportunity to shed new light on the origin of the various order parameters present in this system.

\begin{figure*}[!t]
	\includegraphics[width=7 in]{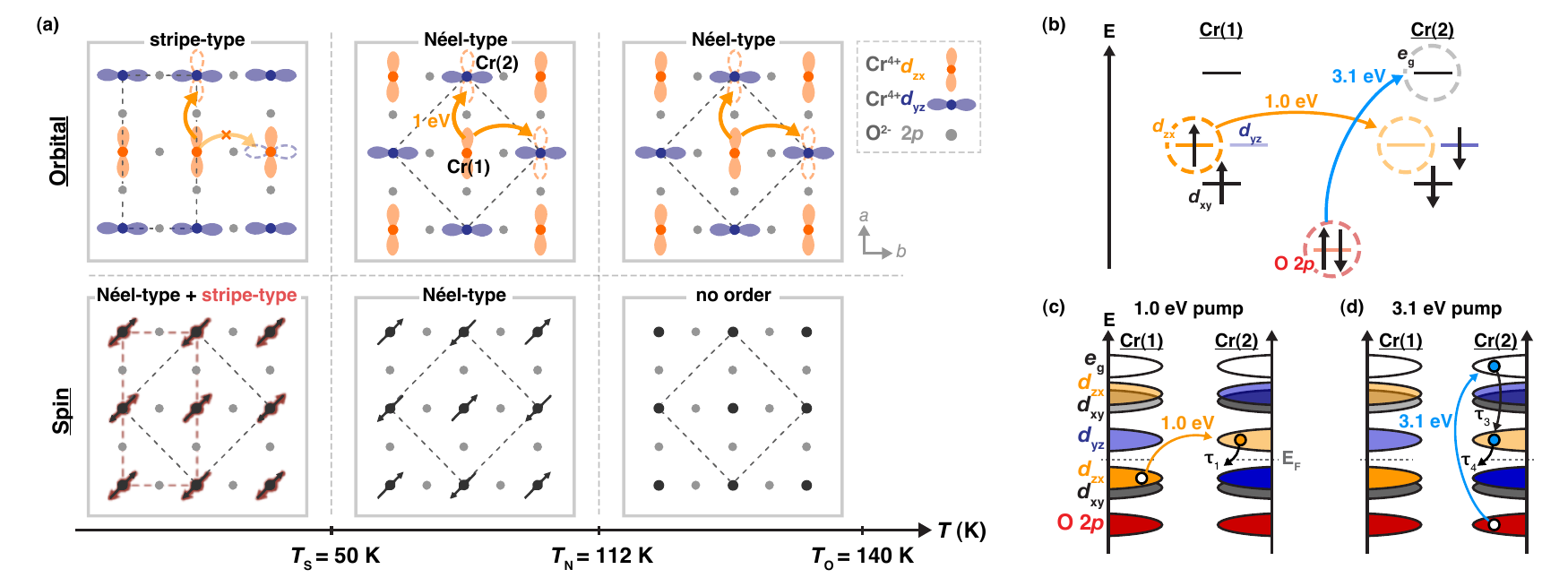}
	\centering
	\caption{(a) Schematic diagrams of the spin- and orbital-ordered ground state in $\alpha$-Sr$_2$CrO$_4$, showing different types of antiferro-type spin ordering and $d_{yz}/d_{zx}$ orbital ordering. We note that the four-lobed shapes of the $d$-orbitals are projected into the $ab$-plane. Arrows indicate possible 1 eV transitions from occupied (filled shapes) to unoccupied (dotted shapes) orbital states in a given ordered phase; transitions from Cr(1) to Cr(2) ions are shown for simplicity, but the reverse can also occur. (b) Energy levels and orbital structures on neighboring sites with Cr(1) and Cr(2) ions for $T \leq T_\textrm{N}$. The $d$-$d$ transitions between $d_{zx}$ orbitals from Cr(1) to Cr(2) (or between $d_{yz}$ orbitals from Cr(2) to Cr(1)) correspond to hopping energies of 1.0 eV. 3.1 eV excitation drives a charge-transfer-type transition from O 2$p$ to Cr $e_g$ orbitals. The relevant electronic structures are displayed along with photoexcitation at (c) 1.0 eV and (d) 3.1 eV. On-site transitions from O $2p$ to Cr $e_g$ states are possible at both Cr(1) and Cr(2) sites.}
	\label{FIG1}
\end{figure*}

 In this work, we thus investigate the non-equilibrium dynamics of spin and orbital order in $\alpha$-Sr$_2$CrO$_4$ by measuring the transient changes in reflectivity at 1.55 eV after femtosecond optical photoexcitation. Importantly, by varying the pump photon energy, we can selectively drive specific microscopic processes that trigger clear $T$-dependent anomalies across the spin and orbital order temperatures. Photoexcitation at 1.0 eV directly perturbs spin and orbital order by inducing spin hopping between neighboring Cr$^{4+}$ ions. In contrast, 3.1 eV photons excite carriers through a charge-transfer-type transition (which does not directly affect spin/orbital order), yet the resulting relaxation dynamics are still influenced by the existing spin and orbital ordered states. Our results thus show that UOS is sensitive to all three of the spin and orbital ordering transitions that have been previously measured via other techniques \cite{Zhu2019,Sakurai2014} through their distinct timescales \cite{Giannetti2016}. Furthermore, we provide new experimental evidence for the existence of a phase transition at $T_\textrm{O}$, with a close relation to the stripe-type spin-orbital order below $T_\textrm{S}$.

 We measured the time-resolved photoinduced changes in reflectivity on an epitaxially grown, 100 nm thick $c$-axis oriented film of $\alpha$-Sr$_2$CrO$_4$, which is usually hard to synthesize in bulk single crystalline form \cite{Zhu2019}. Our experiments were based on a 250 kHz femtosecond (fs) regenerative amplifier producing $\sim$100 fs pulses at 1.55 eV, used to probe the transient response and also to pump an optical parametric amplifier that generated the 1.0 eV pump pulses. A beta-barium borate (BBO) crystal was used to obtain the 3.1 eV pump pulses by frequency doubling the fundamental 1.55 eV pulses. We used a near normal incident geometry for both pump and probe beams, which were linearly cross-polarized along the two equivalent in-plane $a$-axes (since $\alpha$-Sr$_2$CrO$_4$ has tetragonal symmetry). The fluence of the 1.0 eV (3.1 eV) pump pulses was set to be 600 (400) $\mu$J cm\textsuperscript{-2}, generating $\sim$0.1 (0.05) carriers/Cr site and a lattice temperature increase $<$5 K. We measured the time-resolved reflectivity data up to $t = 250$ picoseconds (ps), and verified that the measured response was linear with pump fluence up to $\sim$1 mJ cm\textsuperscript{-2} for both 1.0 eV and 3.1 eV pumping.

 To observe non-equilibrium spin and orbital ordering dynamics, we used our 1.0 eV and 3.1 eV pump pulses to drive specific transitions in the electronic structure of $\alpha$-Sr$_2$CrO$_4$. Two electrons occupy Cr$^{4+}$ $t_{2g}$ states in $\alpha$-Sr$_2$CrO$_4$, leading to a reversed crystal splitting effect that lifts their degeneracy \cite{Ishikawa2017,Jeanneau2019}. This results in a lower energy for the $d_{xy}$ orbital compared to those of the $d_{yz}/d_{zx}$ orbitals (Fig. 1(b)). Theoretical calculations of the ground state orbital structure also revealed an AF pattern in the occupancy of $d_{zx}$ and $d_{yz}$ orbitals on neighboring Cr$^{4+}$ ions (Cr(1) and Cr(2)) below $T_O$, in addition to AF spin ordering below $T_N$ \cite{Ishikawa2017,Pandey2021}. These AF spin and orbital orderings produce different partial densities of states on neighboring Cr ions, resulting in an energy scale of 1.0 eV for inter-site spin hopping from Cr(1) to Cr(2) (Cr(2) to Cr(1)) sites between $d_{zx}$ ($d_{yz}$) orbitals, as indicated by the orange arrows in Fig. 1 \cite{supple,Ishikawa2017}. Additionally, charge transfer-type transitions between neighboring oxygen 2$p$ and Cr $e_g$ orbitals occur at higher energy scales ($\sim$3 eV). We note that these assignments also agree with previous optical conductivity data showing clear peaks at these photon energies \cite{Matsuno2005}. In our experiments, therefore, 1.0 eV photoexcitation directly perturbs spin and orbital order, while 3.1 eV photoexcitation creates carriers whose dynamics are influenced by the existing spin and orbital orders as they relax from higher energy states. We tracked the resulting dynamics with a probe photon energy of 1.55 eV, set below the charge transfer gap energy of 2.0 eV \cite{Matsuno2005} to give general sensitivity to dynamics in all $t_{2g}$ orbitals.

\begin{figure}[!t]
	\includegraphics[width=3.4 in]{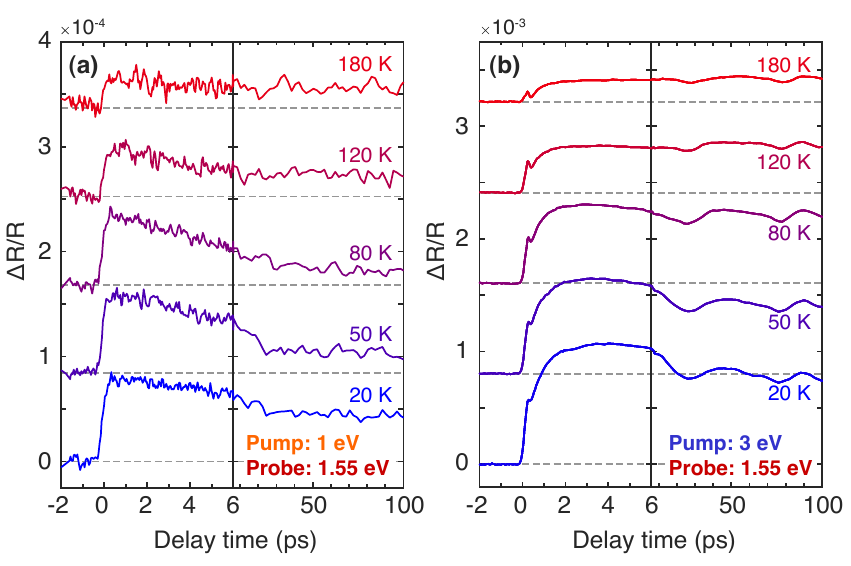}
	\centering
	\caption{Photoinduced time-resolved changes in reflectivity at 1.55 eV after (a) 1.0 eV and (b) 3.1 eV photoexcitation at various temperatures.}
	\label{FIG2}
\end{figure}

 Figure 2 shows $T$-dependent transient reflectivity ($\Delta{R}/R$) data for both pump photon energies, revealing dynamics on both fast (sub-ps) and relatively slow timescales. After 1.0 eV photoexcitation, a step-like feature appears immediately after time $t\sim0$ ps at high temperatures, as exemplified at 180 K, while an additional decaying component arises at low temperatures, as seen at 20 K. In contrast, while the 1.0 eV pump data shows a monotonic decay after $t\sim0$ ps (Fig. 2(a)), the dynamics are more complicated after 3.1 eV photoexcitation, displaying a sharp peak at $t=0.2$ ps 
 as well as a slowly rising component up to $t \sim$ 4 ps (Fig. 2(b)). 
 Finally, there is a slow decay at low temperatures and long time delays after both 1.0 eV and 3.1 eV photoexcitation. We note that the periodic oscillations at a frequency of 22 GHz in the transient reflectivity after 3.1 eV photoexcitation are due to optically driven acoustic phonons \cite{supple,Thomsen1986}; this is well established and will not be discussed further here.

 For more insight, we plotted the $T$-dependence of the $\Delta{R}/R$ signal at different time delays after 1.0 eV pumping in Fig. 3(a). $\Delta{R}/R$ at the peak ($t =$ 0.2 ps) and $t =$ 4 ps clearly deviates from the low-$T$ linear behavior across $T_\textrm{N}$. In contrast, at longer timescales of $t$ = 100 ps, we observe an increase in the $\Delta{R}/R$ signal below the stripe-type ordering temperature ($T_\textrm{S} = 50$ K). This suggests that the dynamics at short and long timescales after 1.0 eV photoexcitation are related to N$\textrm{\'e}$el spin order and stripe-type spin-orbital order, respectively. 
 
 To quantify the timescale for the recovery of magnetic order, we fit our data with a bi-exponential decay function, $\Delta{R(t)}/R = A_1\exp{(-t/\tau_1)}+A_2\exp{(-t/\tau_2)}$, from $t = $ 0.2 ps to 200 ps, as shown in Fig. 3(b). Both the amplitude $A_1$ (Fig. 3(c)) and decay time $\tau_1$ (Fig. 3(d)) of the fast component clearly show a gradual increase below $T_\textrm{N}$. Near the lowest temperatures, $\tau_1\sim10$ ps, revealing the timescale on which AFM order recovers after 1.0 eV excitation. Previous work on antiferromagnetic TM oxides has demonstrated that this time constant is linked to spin-lattice relaxation, i.e. photoexcited electrons rapidly transfer their energy to the lattice, which in turn equilibrates with spins on a timescale given by $\tau_1$ \cite{Bowlan2016,Johnson2015}; this is consistent with our data, and indicates that AFM order restricts carrier hopping below $T_N$ (Fig. 1(a)). We also found that $\tau_2$ exhibits a slower decay time ($\sim$200 ps) above $T_\textrm{S}$ (Fig. S3), likely due to residual lattice heating. However, below $T_\textrm{S}$, $\tau_2$ increases up to $\sim$500 ps, which may be linked to a structural change occurring concurrently with the appearance of stripe-type spin-orbital order (Fig. S6) \cite{supple}.

\begin{figure}[!b]
	\includegraphics[width=3.4 in]{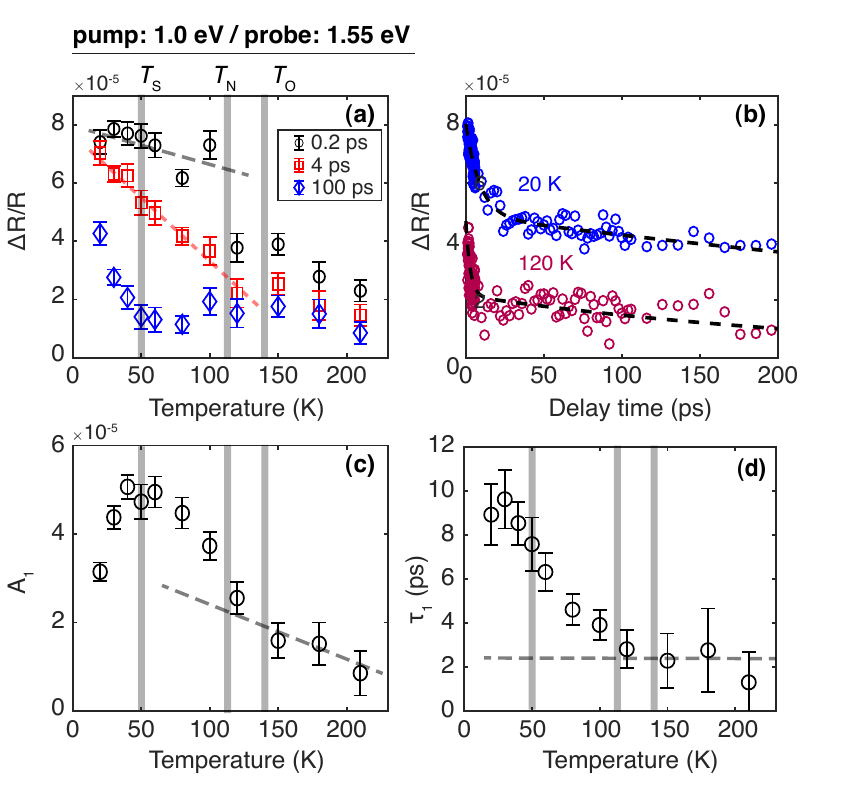}
	\centering
	\caption{Analysis of the temperature-dependent reflectivity transients after 1.0 eV photoexcitation. (a) $\Delta{R}/R$ at time delays of 0.2 ps (black circles), 4 ps (red squares), and 100 ps (blue diamonds). (b) Transient reflectivity data at $T = 20$ K (blue circles) and 120 K (purple circles) along with fits using bi-exponential decay functions (dashed lines). The temperature dependence of the (c) amplitude $A_1$ and (d) decay time $\tau_1$ of the fast component extracted from our fits.}
	\label{FIG3}
\end{figure}

\begin{figure}[!t]
	\includegraphics[width=3.4 in]{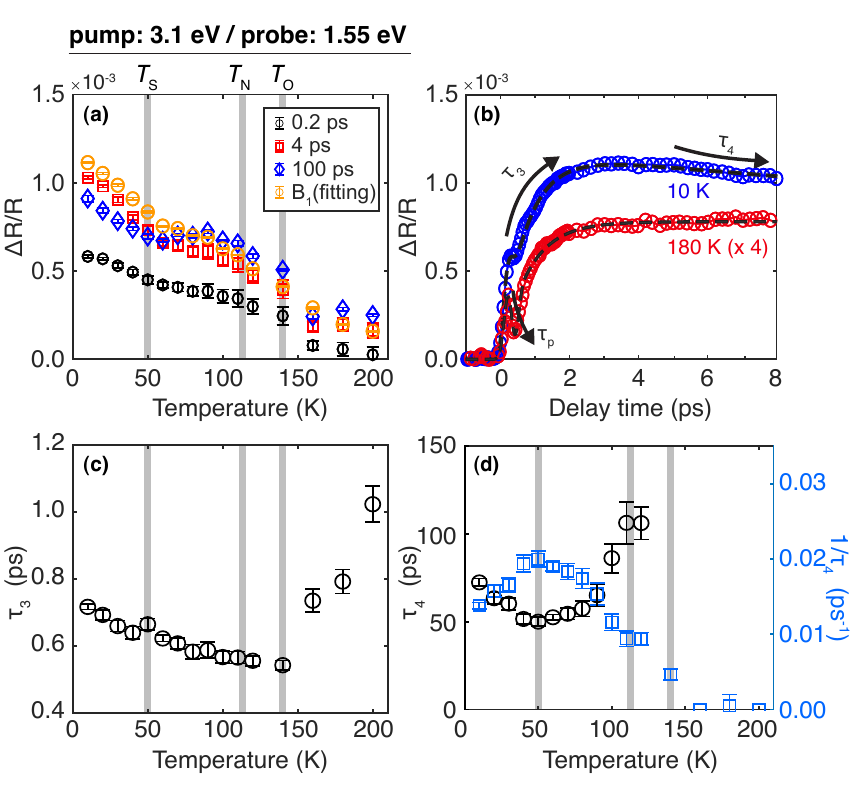}
	\centering
	\caption{Analysis of the temperature-dependent reflectivity transients after 3.1 eV photoexcitation. (a) $\Delta{R}/R$ at time delays of 0.2 ps (black circles), 4 ps (red squares), and 100 ps (blue diamonds). We also plot the amplitude $B_1$ (orange circles) obtained from our curve fits. (b) Transient reflectivity data at $T = 20$ K (blue circles) and 180 K (red circles), along with fits using the model described in the text. The data and fit curve at 180 K are scaled up by a factor of four for clear comparison. (c) The rise time ($\tau_{3}$) and (d) decay time ($\tau_{4}$) obtained from the ultrafast dynamics in (b), revealing a clear anomaly across $T_\textrm{O} = 140$ K.}
	\label{FIG4}
\end{figure}

 In contrast with the transient reflectivity measured after 1.0 eV photoexcitation, which exhibited a relatively simple exponential recovery, the dynamics after 3.1 eV photoexcitation are more complex. Fig. 2(b) reveals three distinct timescales, including (1) the initial sharp peak at $t=0.2$ ps, followed by (2) an increase up to $t\sim4$ ps and (3) a subsequent decay on longer timescales. Fig. 4(a) shows $\Delta{R}/R$ as a function of $T$ at time delays corresponding to these features. While the initial dynamics at $t = 0.2$ ps and 4 ps show only small $T$-dependent anomalies, $\Delta{R}/R$ at 100 ps changes across $T_\textrm{S}$, corresponding to the results after 1.0 eV excitation.
 
 For a more quantitative analysis, we fit the data up to 25 ps with a model function:
 \begin{align}
     \frac{\Delta{R}}{R}(t) = S(t) \times \left(1-\exp{(-\frac{t}{\tau_3})}\right)\nonumber\\
     \times\left(B_1\exp{\frac{-t}{\tau_4}}+C\exp{(\frac{-(t-t_0')}{\tau_p})^2}\right).
 \end{align}
 While a simple two-exponential decay model was sufficient to fit the 1.0 eV data, for the 3.1 eV data we used equation (1), particularly to account for the relatively slow rising dynamics. In this model, $S(t) = (1+erf{(t/\tau_0)})/2$ is a step-function representing the convolution of the material response with the pulse duration of $\tau_0\sim$100 fs. We use $1-\exp{(-t/\tau_3)}$ to fit the slow rising dynamics with a rise time $\tau_3$, and $B \exp{(-t/\tau_4)}$ to fit the subsequent decay with amplitude $B$ and decay time $\tau_4$. The remaining part, $C_1\exp{(-(x-t_0')/\tau_p)^2)}$, is used to fit the initial peak at $t_0' \sim 0.2$ ps. The fitting results at $T =$ 10 and 180 K are displayed in Fig. 4(b) with the raw data, showing that our model can accurately represent the complex dynamics triggered by 3.1 eV pumping.
 
 The initial sharp peak may originate from several different processes, including electron-electron scattering among the photoexcited carriers \cite{Kurz1988} and two-photon absorption; additional measurements are required to conclusively determine its origin. Regardless, the initial charge transfer-type excitation between O 2$p$ to Cr $e_g$ states on neighboring ions driven by 3.1 eV photoexcitation should not directly influence magnetic or orbital order on ultrafast timescales, because neither the initial nor the final state are involved with the spin and orbital order in the Cr $t_{2g}$ states (Fig. 1(d)). 
 
 After the initial peak, the slow rise denoted by $\tau_3$ (Fig. 4(b)) is due to relaxation of the photoinduced carriers from their initial $e_g$ levels into the $t_{2g}$ states examined by our 1.55 eV probe pulse, as indicated by the upper black arrow in Fig. 1(d)). This slow rising component does not appear in the 1.0 eV data (Fig. 2(a)), since our 1.55 eV probe energy is higher than that of the 1.0 eV pump. Interestingly, $\tau_3$ shows a clear anomaly across $T_\textrm{O} = 140$ K, as shown in Fig. 4(c), a transition temperature that has been observed only once via specific heat measurements \cite{Sakurai2014}. This suggests that the orbital configuration changes below $T_\textrm{O}$, which can influence carrier relaxation into the conduction band minima originating from Cr $t_{2g}$ orbitals. 
 
 The subsequent dynamics characterized by $\tau_4$ provide further evidence of a phase transition across $T_\textrm{O}$. This time constant, which is likely due to both on-site and inter-site relaxation (discussed further below), appears below $T_\textrm{O}$ and is characterized by the divergence of its decay time (Fig. 4(d)); we also plot the inverse of $\tau_4$ to make its $T$-dependent anomaly at $T_\textrm{O}$ clear. Additionally, $\tau_4$ increases below $T_\textrm{S}$ (Fig. 4(d)). Therefore, the relaxation dynamics occurring on a timescale of tens of ps after 3.1 eV excitation are directly linked to orbital order via the two distinct phase transitions at $T_\textrm{O}$ and $T_\textrm{S}$.

 For further insight into the orbital configuration, we refer to previous theoretical calculations based on a Kugel-Khomskii model (Fig. 4(a) in \cite{Zhu2019}). Considering the exchange interactions between the neighboring spins and orbitals within $t_{2g}$ states, two different antiferro-type orbital orderings in the $d_{yz}$ and $d_{zx}$ states with N\textrm{\'e}el-type and stripe-type patterns can exist (Fig. 1(a)), depending on the next-nearest-neighbor exchange coupling. These calculations confirmed that the stripe-type orbital order has a lower energy, which should be accompanied by a reconfiguration of the spin degrees of freedom. This agreed with RXS data showing a scattering signal consistent with stripe-type order at the lowest temperatures \cite{Zhu2019}, as well as muon spin rotation data showing a second oscillatory component that may be due to coexisting stripe-type spin order \cite{Sugiyama2014}. These results therefore suggested that stripe-type spin-orbital order is the ground state, but AF-type orbital order can also exist in $\alpha$-Sr$_2$CrO$_4$, depending on inter-site interactions between Cr ions.

 Guided by these theoretical results, our data thus suggests that the pattern of orbital order changes from stripe-type below $T_\textrm{S}= 50$ K to N\textrm{\'e}el-type $d_{yz}/d_{zx}$ orbital order above $T_\textrm{S}$ (Fig. 1(a)), which has been hard to observe due to the lack of clear experimental signatures, except from X-ray scattering measurements \cite{Zhu2019}. For instance, the increase in $\tau_4$ below $T_\textrm{S}$ (Fig. 4(d)) is likely due to the fact that the possible routes for inter-site hopping between the same $d_{zx}$ (or $d_{yz}$) bands at nearest neighbor Cr ions are cut in half when orbital order changes from AF-type to stripe-type. More specifically, as shown in Fig. 1(a), inter-site hopping is feasible above $T_\textrm{S}$ from Cr(1) to both Cr(2) nearest neighbors. However, one of these channels is blocked upon the development of stripe order, causing $\tau_4$ to increase below $T_\textrm{S}$. In addition, we note that some fraction of the carriers photoexcited at 3.1 eV can also relax on-site (contributing to $\tau_4$), but this process should not be directly influenced by the transition to stripe-type order at $T_{S}$. This explains the fact that we do not observe any anomaly across $T_\textrm{S}$ in $\tau_3$ (unlike in $\tau_4$), as this component originates from on-site carrier relaxation from higher lying $e_g$ to $t_{2g}$ states above the Fermi level and is thus not influenced by changes from AF- to stripe-type order.

 The appearance of stripe-type spin-orbital order below $T_\textrm{S}$ also causes the amplitude $A_1$, associated with the spin demagnetization dynamics triggered by 1.0 eV pumping, to be suppressed (Fig. 3(c)). This occurs because the probability of inter-site transitions driven by 1.0 eV photoexcitation between the same spin-polarized $d_{yz}$ (or $d_{zx}$) orbitals at the nearest Cr ions in the stripe-type spin-orbital ordered phase is half of that at higher temperatures, as with $\tau_4$ (Fig. 1(a)). However, the relevant decay time $\tau_1$ for spin relaxation does not change abruptly across $T_\textrm{S}$, suggesting a persistent nature of the N\textrm{\'e}el-type spin order that is stable at all temperatures below $T_\textrm{N}$, though it may be accompanied by stripe-type spin order below $T_\textrm{S}$ \cite{Sugiyama2014}. This corresponds to the previous RXS results, showing a gradual development of the N\textrm{\'e}el-type magnetic Bragg peak across $T_\textrm{S}$ \cite{Zhu2019}.

 Finally, at higher temperatures, the $T$-dependent anomalies in $\tau_3$ and $\tau_4$ across $T_\textrm{O} = 140$ K most likely originate from the phase transition into the AF-type orbital ordered state. Assuming that there is no orbital order above $T_\textrm{O}$, the electronic structure for $T>T_\textrm{O}$ will differ from the orbital-ordered state below $T_\textrm{O}$. This is supported by the fact that the $c$-axis lattice constant in the CrO$_6$ octahedral structure decreases below $T_\textrm{O}$, as shown in Fig. S6 \cite{supple,Jeanneau2019}. Such distortions in the octahedral structure can change the hybridization strength between $t_{2g}$ and $e_g$ orbitals \cite{Landron2008}, which in turn affects the on-site relaxation of photoinduced carriers from $e_g$ states to $t_{2g}$ states after 3.1 eV photoexcitation (Fig. 1(d)). In other words, an increase in $t_{2g}$-$e_g$ hybridization will enable photoexcited electrons in the $e_g$ bands to scatter into the $t_{2g}$ bands more effectively, decreasing $\tau_3$ below $T_\textrm{O}$. Simultaneously, the photoexcited holes will hop from the $O_{2p}$ oxygen states to either Cr(1) or Cr(2) ions, facilitating both on-site and inter-site recombination with the excess electrons in the $t_{2g}$ bands (both processes contribute to $\tau_4$). These considerations thus give us a general picture of carrier dynamics in $\alpha$-Sr$_2$CrO$_4$ (and its sensitivity to orbital order) after 3.1 eV photoexcitation. 

 In conclusion, we used femtosecond optical spectroscopy to track non-equilibrium carrier dynamics in $\alpha$-Sr$_2$CrO$_4$, revealing clear temperature-dependent anomalies across the spin and orbital ordering temperatures. These experimental results are consistent with a theoretical prediction \cite{Zhu2019} of two possible orbital configurations in the $d_{yz}$ and $d_{zx}$ states, i.e. Ne{\'e}l-type order below $T_\textrm{O}$ that turns into stripe-type order below $T_\textrm{S}$. Overall, our results underline the ability of ultrafast optical spectroscopy to distinguish spin and orbital orders through their timescales for coupling to the electronic structure in complex materials.
 
 \section*{Acknowledgement}
 This work was performed at the Center for Integrated Nanotechnologies at Los Alamos National Laboratory (LANL), a U.S. Department of Energy, Office of Basic Energy Sciences user facility, under user proposal 2018BU0083. It was primarily supported through the U.S. Department of Energy, Office of Science, Office of Basic Energy Sciences, Division of Materials Sciences and Engineering via FWP No. 2018LANLBES16 (M.-C.L. and R.P.P.). Work at MIT (C.O., J.L. Z.Z., and R.C.) was supported by the U.S. Department of Energy, Office of Science, Office of Basic Energy Sciences under Contract No. DE-SC0019126.
 
 \section*{Author contributions}
 M.-C.L., R.C., and R.P.P. conceived and designed the project. $\alpha$-Sr$_2$CrO$_4$ single crystalline thin films were grown and characterized by J.L., Z.Z., C.O., and R.C. M.-C.L. performed the experiments with help from N.S.S and L.T.M. The data was analyzed by M.-C.L., C.O., J.L., R.C., and R.P.P. The manuscript was written by M.-C.L. and R.P.P. with significant contributions from C.O., J.L., Z.Z., R.C., and D.A.Y.
  
 \section*{Competing Interests}
 The authors declare no competing interests.

\end{document}